\documentclass{article}            
\usepackage{graphicx}
    
\begin{document}                  

\title{A double supermirror monochromator for neutron instrumentation at LLB}

\author{S. D\'esert[a] \and V. Th\'evenot[a] \and A. Gabriel[a] \and P. Permingeat[a] \and J. Oberdisse[b] \and A. Br\^ulet[a]}

\maketitle 

a Laboratoire Leon Brillouin, CEA Saclay, 91191 Gif-Yvette, France\\
b Laboratoire des Colloides, Verres et Nanomateriaux, 34095 Montpellier, France
                 
\begin{abstract}
The design and characteristics of a double supermirror monochromator for neutron instrumentation at the Laboratoire L\'eon Brillouin is described. The aim of this monochromator is to reduce the intense $\gamma$ radiation produced by conventional velocity selectors and to avoid a direct view of the guide while keeping a comparable neutron transmission (higher than 70\%). The monochromator offers a continuous choice of wavelength selection in the range 0.5 to 2 nm.   
\end{abstract}
\newpage

\section{Introduction}
Neutron scattering spectrometers dedicated to large scale structures studies are generally equipped with velocity selectors. The latter, used as monochromators, select from the cold ``white'' neutron beam a narrow band ($\frac{\Delta \lambda}{\lambda}$ between 10-20\% where $\Delta \lambda$ is the full width at half maximum, FWHM) of neutron velocities by rotating either a helical neutron absorber \cite{dash53} or a series of absorbing discs with slits of helical pitch \cite{rosta91,hammouda92}. Common absorbers are cadmium or gadolinium (Gd) which display very high neutron absorption cross-sections but then produce intense $\gamma$ radiation. 
The latter is a significant problem when using image plate detectors sensitive to neutrons \cite{wilkinson92}, since their detection principle relies upon conversion of neutron to secondary particles which cannot be discriminated from ambient gamma radiation present in a reactor environment.\\
For our very small angle neutron scattering (VSANS) spectrometer \cite{desert06}, we have replaced the velocity selector with a double supermirror monochromator. Its principle is that of a classical double crystal monochromator but with specific supermirror (SM) elements instead of single crystals. In this paper, we report on the design, characteristics and results of this original double SM monochromator, now installed at the end of the bender G5-bis at LLB.

\section{Supermirror monochromator}
\subsection{Constraints for the VSANS monochromator}
The monochromator device of the VSANS spectrometer at LLB has to deliver long wavelengths $\lambda$ with $\frac{\Delta \lambda}{\lambda}$ of 10 to 15\% in order to access small q values while having a reasonable neutron flux. In addition, its transmission has to be high, around 0.7. Due to the $\gamma$ senstivity of the image plate detector on the spectrometer, it is not possible to use conventional velocity selectors because of the intense $\gamma$ radiation emitted from Gd absorber. Among possible absorbing materials that produce less intense $\gamma$ emission, we find isotopic lithium, $^{6}$Li, which is commercially available. Unfortunately, its absorption cross-section is 52 times smaller than that of Gd \cite{sears92}. Therefore, a new design of the velocity selector mechanics with thicker $^{6}$Li discs remains to be manufactured \cite{dohmer00}. Another possible monochromator is a double crystal monochromator of large mosaicity able to select variable long wavelengths and maintain the outgoing monochromatic beam parallel to the incident beam when changing the wavelength. Only a few materials with suitable lattice spacing ($d$) and mosaicity can possibly be used. Graphite intercalation compounds are the most interesting materials for this purpose but the maximum selected wavelength is quite small, lower than 0.89 nm \cite{sears97}. Moreover, the crystal mosaicity obtained (1 or 2$^{\circ}$) leads to a very good wavelength resolution but with a low bandwidth, i.e. low integrated flux. Such a resolution is not required for SANS whose aim is rather to enhance flux at the sample. The use of crystal monochromators is also restricted by high-order reflections which are particularly important at long wavelengths.\\ 
An alternative to the crystal monochromator is the multilayer thin film monochromator \cite{saxena77}. The multilayers are obtained by alternately depositing thin films of two materials on a substrate (glass, silicon, ...) upon which the neutron beam is diffracted according to Bragg's law. Detailed studies of multilayers \cite{saxena77,mezei77} have shown that they can be used as efficient monochromators. In addition, the diffracted monochromatic beam has a weak higher order contamination. From a practical point of view, since the d-spacing of such monochromators can be chosen in a wide range of thicknesses (from 2 nm up to 100 nm) and their distribution $\Delta d$ can be varied during manufacture, it is possible to choose the critical angle $\theta_{c}$ and the bandwidth, $\Delta \theta_c$, of the multilayer film. Such multilayers are already used as monochromator guides at LLB\footnote{http://www-llb.cea.fr/spectros/spectros.html} (for example on G5-5 to deliver 0.8 nm wavelength with $\frac{\Delta \lambda}{\lambda}$=0.1, and on G2-1 to deliver 0.4 nm and $\frac{\Delta \lambda}{\lambda}$=0.05). In the past decades, considerable work on neutron supermirrors have improved both the limit of critical angle for neutron reflection and the transmission of short wavelengths. Ni/Ti supermirrors \cite{elsenhans94} are used in numerous neutron research centres as classical neutron guides, now often up to 3 times the critical angle of ordinary Ni.\\
Here, we use them as a monochromator. In order to keep the scattering angle constant when varying the selected wavelength, we designed a double SM monochromator with Ni/Ti supermirrors of bandwith $\frac{\Delta \theta}{\theta}$ around 15\% at the critical angle 3$\theta_c$. The reflectivity at 3$\theta_c$ is high, $R\approx 0.8$. The device will allow us to change the wavelength in the range 0.5 to 2 nm.

\subsection{Design}
\begin{figure}
\caption{Drawing of the double SM monochromator installed at the end of the bender G5-bis at LLB.}
\includegraphics[width=8cm]{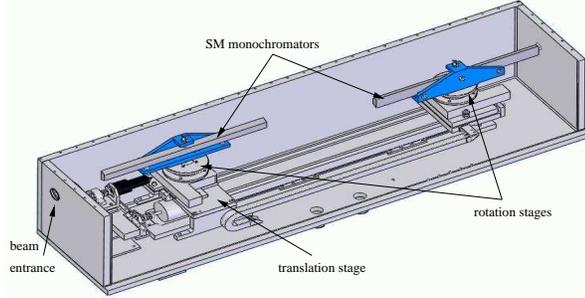}
\label{schemas}
\end{figure}

The double reflection SM monochromator is made of two Ni/Ti SM monochromators (purchased from Swissneutronics and mounted by CILAS) with a critical angle  3$\theta_c$ and a $\frac{\Delta \theta}{\theta}$ of 15\%. Each element is 30 mm high and 60 cm long and made of two 30 cm long pieces glued on a glass support. The two mirrors are mounted on rotation stages and the second one is also mounted on a 80 cm translation stage parallel to the incident beam direction (see figure \ref{schemas}). Both mirrors should always be at the same rotation angle relative to the incoming beam. The translation therefore enables the outgoing monochromated beam to remain fixed while varying the wavelength. A 55 mm offset relative to the incident beam prevents the detector from viewing the primary guide and will therefore decrease the $\gamma$ background measured by the image plate. The 2 m long monochromator is placed under vacuum. With this setup and the characteristics of the SM monochromators, the wavelength can be continuously varied from 0.5 up to 2 nm, corresponding to rotation angles from 1.5 to 6$^{\circ}$.

\subsection{Reflectivity measurements}
\label{refmesures}
\begin{figure}
\caption{Reflectivity curves for two supermirror monochromators 3$\theta_c$ measured on EROS at LLB.}
\includegraphics{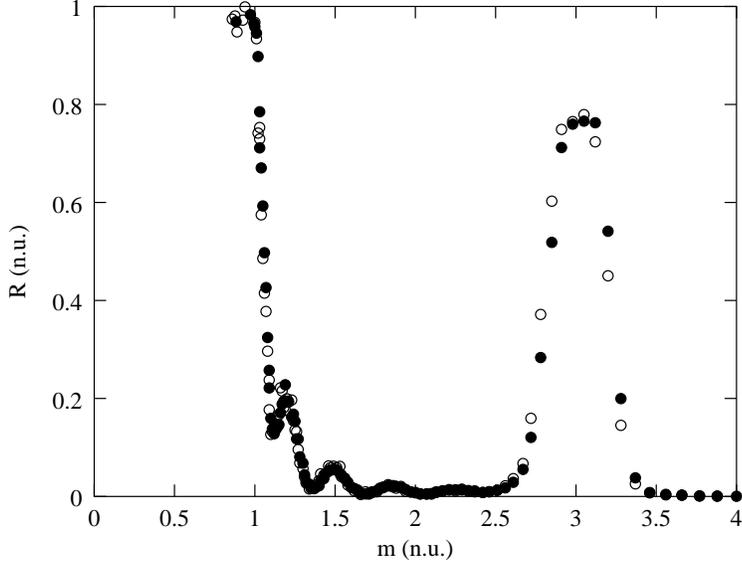}
\label{monokR1}
\end{figure}

\begin{figure}
\caption{Reflectivity curves measured after a single reflection on the SM monochromator ($\circ$) and calculated by squaring the previous curve assuming a second reflection on an identical SM element ($\bullet$). $\frac{\Delta m}{m}$ is reduced from 14 to 11\%. The maximum reflectivity at $m$=3 after the second reflection is 0.7.}
\includegraphics{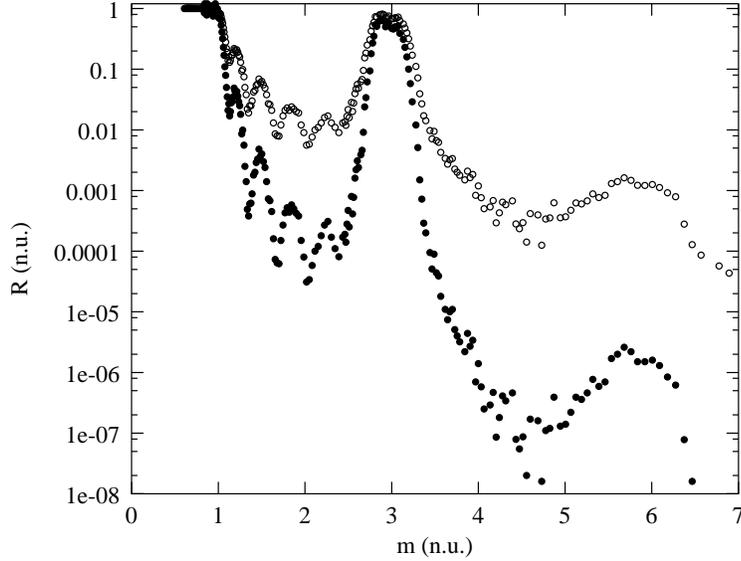}
\label{monokR1R2}
\end{figure}

Reflectivities of the 60 cm long SM monochromators were measured on the EROS reflectometer \cite{menelle03} at LLB. Recall that for a given material, total neutron reflection (R=1) occurs for all grazing incident angles $\theta$ lower than a critical angle $\theta_c(\lambda)=\lambda \sqrt{\frac{Nb}{\pi}}$ where $N$ stands for the number density of atoms and $b$ is the coherent scattering length of the material. For natural Ni, $Nb$=9$\times$10$^{-4}$nm$^{-2}$ \cite{sears92}.\\
Figure \ref{monokR1} shows the reflectivity curves of two SM plotted as a function of $m$ given by:
\begin{equation}
m=\frac{\theta}{\theta_c \lambda}
\label{m}
\end{equation}
with $\theta_c$=0.991$^{\circ}$\AA$^{-1}$for Ni. $\frac{\Delta m}{m}$ is around 0.14 and the maximum peak reflectivity measured at $m$=3 is very high, around 0.85.
Oscillations can be observed between m=1 and m=3. These contaminations of the reflected monochromated beam have to be studied in detail. However, since double reflection is used, the relative importance of these contaminations with respect to the selected monochromatic beam is significantly reduced. Figure \ref{monokR1R2} shows that the weak oscillating contributions are largely attenuated after the second reflection whereas the total reflection plateau remains equal to unity. These effects will be discussed in section \ref{doublereflec}. Another effect of the second reflection is to increase the wavelength resolution, $\frac{\Delta m}{m}$ reduces to 11\% instead of 14\% for a single reflection. Finally, the peak intensity of the reflected beam is also reduced to 0.7 (=0.85$^2$).
The useful neutrons for a small angle neutron scattering spectrometer have very small divergence; the maximum divergence of useful neutrons is +/- 0.1$^{\circ}$ in the case of our VSANS. Therefore we will hereafter consider the transmission of the monochromator to be the same as the reflectivity. In the case of experiments requiring larger divergence, the transmission of useful neutrons will decrease because of the geometrical requirement that if the neutrons don't intercept both mirrors then they will not be transmitted by the monochromator. In order to obtain the real transmission of our setup, a comparison of ToF measurements with the white incoming beam is to be performed as a future work.

\section{Double reflection supermirror monochromator}
\subsection{Time of flight (TOF) measurements}
\begin{figure}
\caption{Time of flight measurement at 5.4 ($\circ$), 8 ($\bullet)$ and 12 \AA \hspace{0.3mm}($\diamond$). The intensities have been rescaled to unity for convenience and the dashed line represent the best gaussian fit of the data.}
\includegraphics{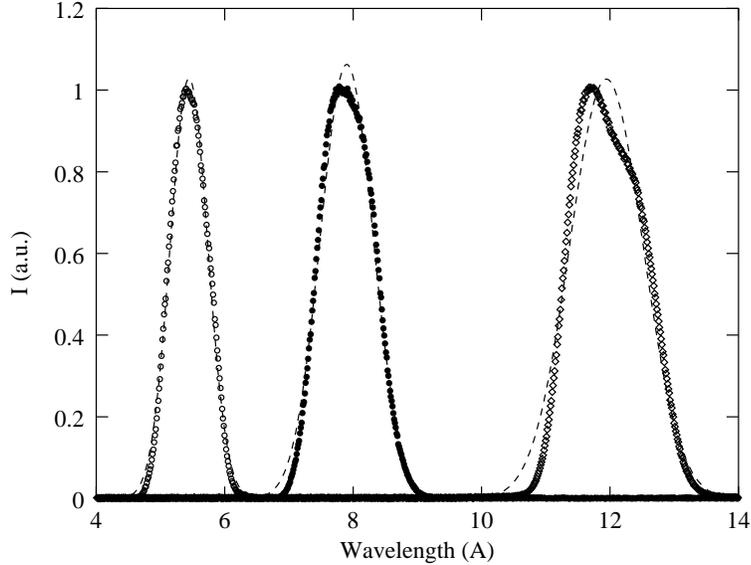}
\label{TOF}
\end{figure}

For time of flight mleasurements (TOF), a disc chopper with rotational frequency of 50 Hz is located at the exit of the double SM monochromator. The neutron beam is defined by a 2 mm diameter aperture placed at the front of the chopper and a 5 mm wide slit cut in the chopper at a radius of 23 cm. Neutrons are detected by a simple $^3$He counter located at 3.22 m from the chopper. Neutron pulses are counted with a multichannel scaler triggered with a photosensor, once per chopper rotation.
Pulse spreading due to the finite aperture of the TOF device corresponds to an increase of 0.1 \AA \hspace{0.3mm} in the monochromated beam bandwidth.
Figure \ref{TOF} presents the measurements at three different wavelengths with their best fit to gaussian functions and the fitted parameters are listed in table \ref{TableTOF}. After subtraction of the spreading due to the chopper aperture, $\frac{\Delta \lambda}{\lambda}$ is close to 11\% for each wavelength.

\subsection{Wavelength contamination}
\label{doublereflec}
\begin{figure}
\caption{Wavelengths distribution after the monochromator set for 6 \AA \hspace{0.3mm} neutrons. TOF measurements with a large aperture to enhance the oscillations ($\bullet$) and I($\lambda$)R$^{2}$($\lambda$) computation ($\circ$) where I($\lambda$) is the 'white' beam distribution and R($\lambda$) is the measured reflectivity of figure \ref{monokR1}.}
\includegraphics{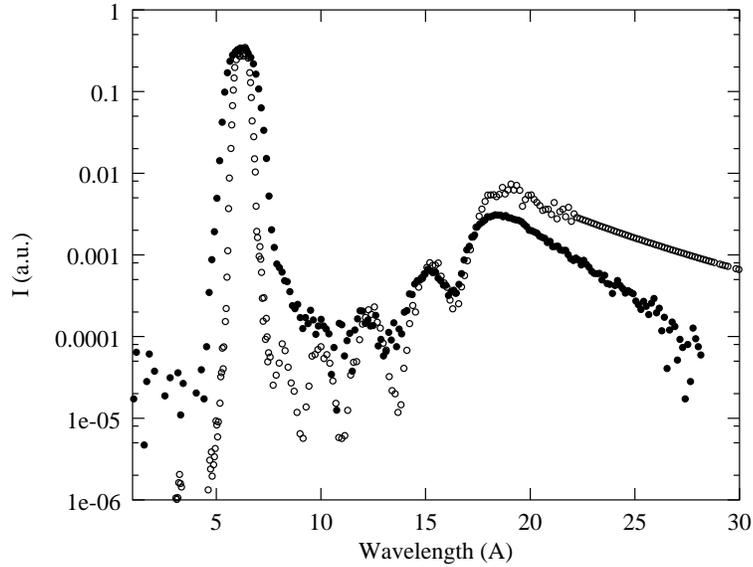}
\label{monokR2}
\end{figure}

\begin{figure}
\caption{Bandpass filter location inside the monochromator.}
\includegraphics{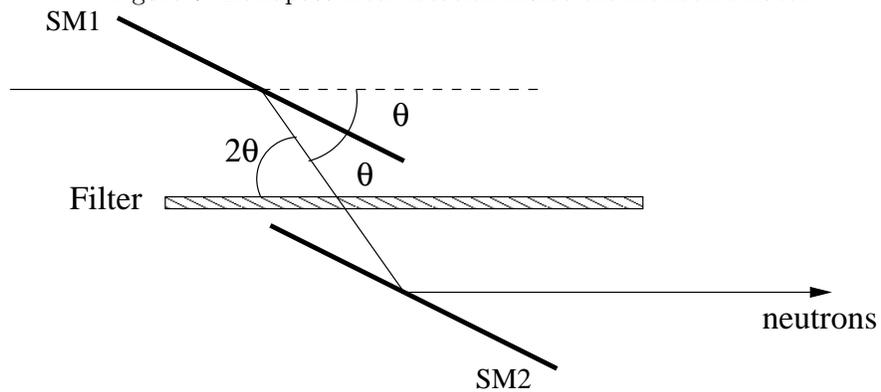}
\label{filtre}
\end{figure}

\begin{figure}
\caption{Wavelength distribution on logarithmic scale (monochromator set for 7 \AA \hspace{0.3mm} neutrons) without ($\circ$) and  with (+) the bandpass filter inserted inside the monochromator.}
\includegraphics{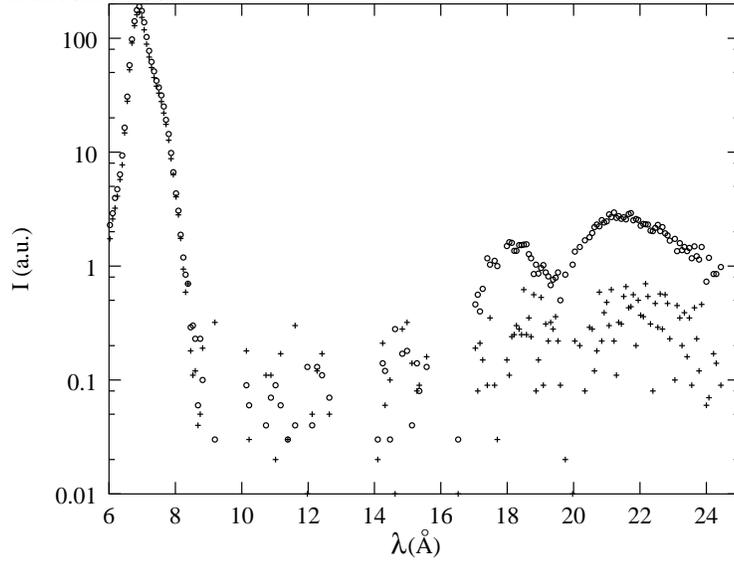}
\label{manipFiltre}
\end{figure}

As mentioned in section \ref{refmesures}, the reflectivity plateau could give rise to unwanted wavelengths exiting the monochromator. Figure \ref{monokR2} shows the measured and calculated contaminations for $\lambda$=6 \AA. The flux distribution calculated after the monochromator is the product of the reflectivity and the incoming flux distribution. This figure shows that intermediate and long wavelengths (higher than 17 \AA) are being transmitted by the monochromator with respectively negligible (lower than 10$^{-3}$) and weak (up to 10$^{-2}$) relative intensity compared to the peak wavelength intensity.\\Such an effect is due to the large wavelength distribution of the cold source.
Our spectrometer characteristics will not be affected by this contamination since it is designed to obtain its best performance for 15 \AA \hspace{0.3mm} neutrons. In this condition, 45 \AA \hspace{0.3mm} and higher unwanted neutrons have negligible flux. Nevertheless, in order to select shorter wavelengths, a neutron lowpass filter \cite{anderson96} can be added inside the monochromator. We have checked this solution with a m=3.6 SM filter deposited on a thin silicon wafer (courtesy of Swissneutronics) parallel to the incoming beam and located between the SM monochromators as shown in figure \ref{filtre}. Unwanted long wavelengths are reflected while the short wavelengths are transmitted. In this setup, the filter remains fixed because its angle relative to the beam will always be twice the SM angle.
Figure \ref{manipFiltre} shows TOF measurements for a selected wavelength of 7 \AA \hspace{0.3mm} with and without the filter. The selected wavelength intensity is reduced to 90\% of the intensity without the filter. This loss corresponds to the transmission through the silicon wafer. The contamination is reduced by roughly a factor 6 which is consistent with the filter characteristics of 85\% transmission meaning that only 15\% of the contamination passes the filter. Note that it is possible to use a better filter, with 95\% transmission, which would thus reduce the contamination by a factor of 20.
\subsection{Tuning the wavelength bandwidth}
\begin{figure}
\caption{Attenuation factor at 6 ($\bullet$) and 15 \AA \hspace{0.3mm} ($\circ$), on logarithmic scale, (using equation \ref{TuningGeneral} with $\frac{\Delta \theta_1}{\theta1}$=$\frac{\Delta \theta_2}{\theta_2}=0.15$) as a function of $\frac{\Delta \bar{\theta}}{\bar{\theta}}$.}
\includegraphics{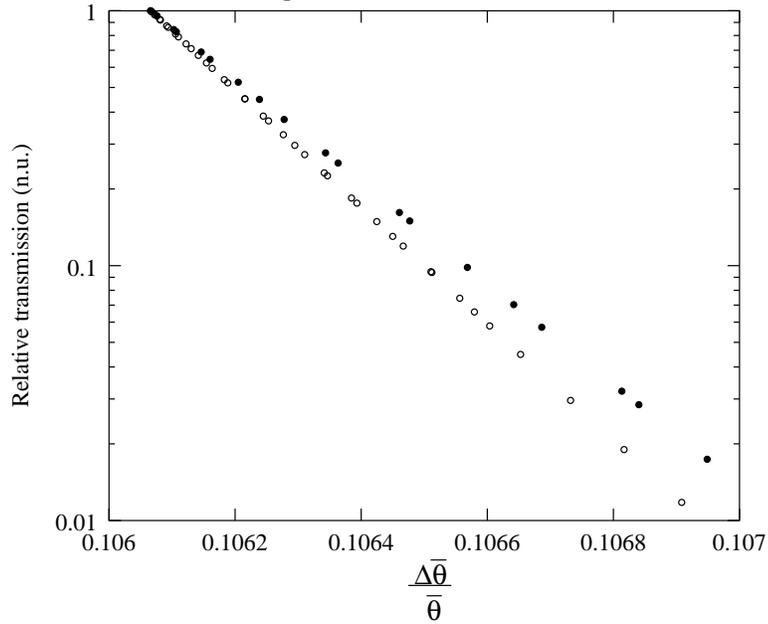}
\label{tuning}
\end{figure}

It is theoretically possible to change the wavelength bandwidth while keeping the reflected beam at a constant position by choosing different angles for each mirror. Let us consider the simplified case of two SM monochromators of the same critical angle $m \theta_c$ (our case, $m$=3). Assuming that the peak reflectivity at $m$=3 behaves like a gaussian function with maximum reflectivity of unity, then the wavelength intensity distribution $I(\theta)$ after a double reflection is given by the product:\\
\begin{equation}
\label{EqTuning}
I(\theta)  =  \exp \left [- \frac{(\theta - \theta_1)^2}{2 \sigma_{\theta_1}^2} \right ] \times \exp \left [-\frac{(\theta - \theta_2)^2}{2 \sigma_{\theta_2}^2} \right ]
\end{equation}
where $\theta_i$ is the rotation of mirror $i$, $\sigma_{\theta_i}$ is the standard deviation corresponding to the neutrons with incoming angle $\theta_i$ and $\sigma_{\theta} = \frac{\Delta \theta}{2\sqrt{2 \ln 2}}$. Let us define the ratio $\delta = \left ( \frac{\sigma_{\theta_1}}{\sigma_{\theta_2}} \right )^2$ so that $\delta = \left ( \frac{\theta_1}{\theta_2} \right )^2 \left (\frac{\frac{\Delta \theta_1}{\theta_1}}{\frac{\Delta \theta_2}{\theta_2}} \right )^2$.\\
Introducing $\delta$ in equation \ref{EqTuning} leads to the product of a gaussian function and a $\theta$ independent factor:
\begin{equation}
\label{EqTuning1}
I(\theta) =  \exp \left [- \frac{\left (\theta - \frac{\theta_1 + \delta \theta_2}{1+ \delta}\right )^2}{2 \left (\frac{\sigma_{\theta_1}}{\sqrt{1+\delta}} \right )^2}\right ] \times \exp \left [- \frac{\delta}{2(1+\delta)\sigma_{\theta_1}^2}(\theta_1 - \theta_2)^2 \right ]
\end{equation}
The first exponential term shows that the angle corresponding to the maximum intensity is $\bar{\theta}= \frac{\theta_1 + \delta \theta_2}{1+\delta}$ and its associated standard deviation $\sigma_{\bar{\theta}}=\frac{\sigma_{\theta_1}}{\sqrt{(1+\delta)}}$. The second term of equation \ref{EqTuning1} corresponds to an attenuation factor, which at best equals one when $\theta_1=\theta_2$. Equation \ref{EqTuning1} means that changing the bandwidth of the monochromated beam produces a decrease in the transmission. The resulting wavelength resolution, $\frac{\Delta \bar{\theta}}{\bar{\theta}}$ is given by:
\begin{equation}
\label{TuningGeneral}
\frac{\Delta \bar{\theta}}{\bar{\theta}}=\frac{\Delta \theta_1}{\theta_1}\times \frac{\left [\left (\frac{\Delta \theta_1}{\theta1}\right )^2 \theta_1^2+\left (\frac{\Delta \theta_2}{\theta_2}\right )^2 \theta_2^2 \right ]^{\frac{1}{2}}}{\left (\frac{\Delta \theta_1}{\theta1}\right )^2 \theta_1+\left ( \frac{\Delta \theta_2}{\theta_2}\right )^2\theta_2}.
\end{equation}
$\frac{\Delta \bar{\theta}}{\bar{\theta}}$ is minimum when $\theta_1$=$\theta_2$ and thus can only be increased. Figure \ref{tuning} presents the intensity attenuation as a function of $\frac{\Delta \bar{\theta}}{\bar{\theta}}$ at two wavelengths. The transmitted intensity decreases very rapidly for small change in the resolution. Obviously, varying the monochromator SM angle to change the wavelength resolution is not useful.\\
Setting $\theta_2$= $\theta_1$ in equation \ref{TuningGeneral} leads to:
\begin{equation}
\label{T}
\frac{\Delta \bar{\theta}}{\bar{\theta}}=\frac{\Delta \theta_1}{\theta_1}\times \frac{\frac{\Delta \theta_2}{\theta_2}}{\sqrt{\left (\frac{\Delta \theta_1}{\theta1}\right )^2+\left (\frac{\Delta \theta_2}{\theta_2}\right )^2}}.
\end{equation}
In our case, $\frac{\Delta \theta_1}{\theta_1}$=$\frac{\Delta \theta_2}{\theta_2}$ leads to the minimum wavelength resolution of $\frac{\Delta \theta_1}{\sqrt{2}\theta_1}$. However, according to equation \ref{T}, using SM monochromators with different $\frac{\Delta \theta}{\theta}$, it is possible to obtain monochromators with other wavelength resolutions while maintaining an outgoing beam parrallel to the incident beam.

\section{Conclusion}
The double reflection SM monochromator is one application resulting from developments of SM with high critical angles. Such a monochromator is now used at LLB for the VSANS instrument. It allows a continuous selection of long wavelengths between 0.5 and 2 nm with a transmission of 70\%. The wavelength resolution $\frac{\Delta \lambda}{\lambda}$ chosen for the VSANS instrument is 11\%. The main advantages of such a monochromator are to reduce the $\gamma$ emission from neutron absorption occuring with conventional velocity selectors, and avoid a direct view of the guide while keeping competitive performance. Therefore, it is a good alternative to velocity selectors.
A possible extension of this double SM technique is to use polarizing SM to allow measurements with polarized neutrons. Another improvement would consist of adding horizontal SM monochromators in order to obtain a two-dimensional monochromator.  

The authors wish to thank F. Coneggo and P. Lambert for the electronic devices and A. Menelle for the reflectivity measurements.

\bibliography{sas1}
\bibliographystyle{plain}

\begin{table}
\begin{center}
\begin{tabular}{|c|c|c|c|}
$\lambda$ (\AA) & $\sigma_{\lambda}$ (\AA) & $\Delta \lambda ^{*}$ (\AA) & $\frac{\Delta \lambda}{\lambda}$ (\%)\\
5.44 & 0.292 & 0.588 & 10.8\\
7.90 & 0.411 & 0.868 & 11.0\\
11.9 & 0.571 & 1.245 & 10.4\\
\end{tabular}
\end{center}
\caption{\textnormal{Mean wavelength, standard deviation and resulting $\frac{\Delta \lambda}{\lambda}$ for gaussian fits of figure \ref{TOF}. $\Delta \lambda ^{*}$ is the adjusted bandwidth corrected by half the experimental wavelength spread (0.1 \AA).}}
\label{TableTOF}
\end{table}

\end{document}